\documentclass[5p]{elsarticle1} 

\usepackage{amsmath}

\usepackage{color}

\newcommand{\be}{\begin{equation}}
\newcommand{\bel}[1]{\begin{equation}\label{#1}}
\newcommand{\ee}{\end{equation}}
\newcommand{\refe}[1]{\unskip~(\ref{#1})}

\begin{document}

\begin{frontmatter}

\title{A differentiation formula, with application to the two-dimensional Schr\"odinger equation}

\author{Alexander Pikovski}
\address{JILA, NIST \& Department of Physics, University of Colorado, Boulder, CO 80309, USA
}



\begin{abstract}
A method for obtaining discretization formulas for the derivatives of a function is presented,
which relies on a generalization of divided differences. These modified divided differences
essentially correspond to a change of the dependent variable. 
This method is applied to the numerical solution
of the eigenvalue problem for the two-dimensional Schr\"odinger equation, where standard methods
converge very slowly while the approach proposed here gives accurate results.
\end{abstract}


\end{frontmatter}



\section{Introduction}

A discrete representation of the derivative of a function is often required 
in numerical computations. In order to solve an equation involving derivatives,
one introduces a grid (mesh) and approximates the original equation by a system of
equations which use the values of the unknown function on the grid points. 
Such a method can be used to solve, for example, a two-point boundary-value problem
or an eigenvalue problem involving an ordinary differential equation.

Numerical methods which discretize derivatives on a grid rely on the fact that 
the discretization error decreases as the grid spacing is decreased.
However, the accuracy of discretization
can vary considerably. If the unknown function changes rapidly in a short interval,
usually many grid points are required in order to capture its behavior sufficiently well in this region. 
In some cases, it is not practical to decrease the grid spacing until convergence is reached
because the rate of convergence is too slow. Instead, one has either to re-formulate the problem to 
make it more well-behaved or to adapt the numerical method to the particular problem.
In adapting the numerical method to the problem, one may proceed along two directions.
One may construct a grid which better approximates the unknown function~\cite{Roos,Manteuffel}, 
e.g.\ which is denser in the region where the function changes rapidly.
Another way is to keep a generic grid, but to adapt the expressions which are used
to discretize the derivatives in the particular problem; this has been often discussed in the context of boudary layers~\cite{Roos,Zadorin}.

In this paper, we present a conceptually simple method to discretize derivatives of a function
which is close to another function.
The presented results can be viewed as a generalization of a formula of Zadorin~\cite{Zadorin}.

The method presented here uses modified divided differences.
The subject of divided differences is classical and well-known.
Here, after a short presentation of divided differences (Sec.\ \ref{sec-2}), we discuss modified
divided differences (Sec. \ref{sec-3}) and give the differentiation formulas.
These are applied to the numerical solution for the two-dimensional Schr\"odinger equation  
and compared to other techniques (Sec. \ref{sec-4}), showing that the present method is superior.

\section{Divided differences and differentiation formulas}
\label{sec-2}

The divided differences (of order $1, 2, \ldots$) of a function $f$ are defined by~\cite{KK,Hildebrand}:
\begin{align}
f(x_0,x_1) &= \frac{f(x_1) - f(x_0)}{x_1 - x_0},\\
f(x_0,x_1,x_2) &= \frac{f(x_1,x_2) - f(x_0, x_1)}{x_2 - x_0},
\quad
\ldots
\end{align}
They can be shown to be symmetric in their arguments. Another property is that the $k$-th order
divided differences of a polynomial of degree $n$ result in a polynomial of degree $n-k$
(and zero if $k>n$). This is seen~\cite{KK,Hildebrand} by considering the first divided difference of $x^n$ at points $x$ and $a$, 
which is
\bel{dd-1}
\frac{x^n-a^n}{x-a} = x^{n-1} + x^{n-2}a + \ldots + a^{n-1} .
\ee

For any function $f$ there is the exact identity known as Newton's formula:
\begin{multline}\label{NF}
\!\! f(x) \!=\! f(a_0) + (x-a_0) f(a_0,a_1) + (x-a_0) (x-a_1) f(a_0,a_1,a_2)  \\[2pt] 
\, + \ldots 
+ (x-a_0) \cdots (x-a_{n-1}) f(a_0, \ldots, a_n) 
+ E(x)
\end{multline}
with
\begin{align}\label{NFE1}
E(x) &= (x-a_0) \cdots (x-a_n) f(a_0, \ldots, a_n, x) \\
& = (x-a_0) \cdots (x-a_n) \frac{f^{n+1}(\xi)}{(n+1)!} \label{NFE2}
\end{align}
where $\xi$ is some number lying between the largest and the smallest of the arguments $(a_0, \ldots, a_n, x)$.
If the error term $E(x)$ is omitted from Eq.~\refe{NF}, one obtains Newton's interpolating polynomial
(of order $n$) for the function $f$ at the points $a_0, \ldots, a_n$.

Expressions for the discretized derivative can be deduced from Newton's interpolation formula by 
differentiating and, if needed, by further manipulations~\cite{KK}. Here we only consider the following 
two formulas.
Considering Eq.~\refe{NF} up to the linear term (i.e.\ with higher terms omitted), 
differentiating it with respect to $x$, and putting $a_0=x$, $a_1=x+h$ gives
\bel{f1}
f'(x) \approx f(x,x+h) = \frac{f(x+h)-f(x)}{h} .
\ee
The same procedure up to quadratic terms and differentiating twice 
with $a_0=x-g$, $a_1=x$, $a_2=x+h$ gives
\bel{f2}
f''(x) \approx 2 f(x-g,x,x+h) .
\ee
It is helpful to know that the second divided differences can also be written in the form
\begin{multline}\label{f2-uneq}
 f(a_0,a_1,a_2) = 
\frac{f(a_0)}{(a_0-a_1)(a_0-a_2)} \\+
\frac{f(a_1)}{(a_1-a_0)(a_1-a_2)} +
\frac{f(a_2)}{(a_2-a_0)(a_2-a_1)} .
\end{multline} 
For $g=h$ Eq.~\refe{f2} gives the well-known
\bel{f2uniform}
f''(x) \approx \frac{f(x-h)-2f(x)+f(x+h)}{h^2} .
\ee

The differentiation formulas of Eqs.~\refe{f1} and \refe{f2} are exact for polynomials up to degree
one and two, respectively, as follows from the property of divided differences of polynomials mentioned above.
The error term of these approximations to the derivatives can be found from Newton's formula,
differentiating the error term of Eq.~\refe{NFE1} or \refe{NFE2}. It is seen that the
error of the approximation \refe{f2} is proportional to $f'(\xi)$ and $f''(\xi)$ evaluated at
some points $\xi$ in the interval $x-g < \xi < x+h$.

\section{Modified divided differences and differentiation formulas}
\label{sec-3}

Modified divided differences with respect to a function $\Phi$ will now be introduced.
The function $\Phi(x)$ is assumed to be strictly increasing.
Define the modified divided differences with respect $\Phi(x)$ by
\begin{align}
f_\Phi(x_0,x_1) &= \frac{f(x_1) - f(x_0)}{\Phi(x_1) - \Phi(x_0)},\\
f_\Phi(x_0,x_1,x_2) &= \frac{f_\Phi(x_1,x_2) - f_\Phi(x_0, x_1)}{\Phi(x_2) - \Phi(x_0)},\quad
\ldots
\end{align}
Clearly, for $\Phi(x)=x$ they reduce to the ordinary divided differences. We also have
\be
f_\Phi(x_0, \ldots x_n) = (f\circ\Phi^{-1})(\Phi(x_0), \ldots \Phi(x_n)) ,
\ee
noting that the inverse function $\Phi^{-1}$ exists since $\Phi(x)$ is strictly increasing.
Therefore, the properties of the modified divided differences can be deduced from the properties of the ordinary 
divided differences. The $k$-th modified divided differences of a polynomial of degree $n$ in the variable $y=\Phi(x)$
give a polynomial in $y$ of order $n-k$, since the equivalent of Eq.~\refe{dd-1} is 
$\frac{\Phi(x)^n-\Phi(a)^n}{\Phi(x)-\Phi(a)}$.
Newton's formula for the modified divided differences reads
\begin{multline}\label{NF-mod}
f(x) = f(a_0) + (\Phi(x)-\Phi(a_0)) f_\Phi(a_0,a_1) + \\
(\Phi(x)-\Phi(a_0)) (\Phi(x)-\Phi(a_1)) f_\Phi(a_0,a_1,a_2) + \ldots
\end{multline}
and the higher terms and the error term (omitted here for brevity) are constructed 
as in Eqs.~\refe{NF}, \refe{NFE1}.

The two-point formula for the approximate derivative is obtained by 
differentiating Eq.~\refe{NF-mod} up to the linear term and 
putting $a_0=x$, $a_1=x+h$:
\bel{mf1}
f'(x) \approx \Phi'(x) f_\Phi(x,x+h) = \Phi'(x) \frac{f(x+h)-f(x)}{\Phi(x+h)-\Phi(x)} .
\ee
This formula [Eq.~\refe{mf1}] as well as the linear term of Eq.~\refe{NF-mod} were suggested 
by Zadorin~\cite{Zadorin}.
Differentiating Eq.~\refe{NF-mod} twice, up to the three-point term, and putting $a_1=x$ gives
\begin{align}
f''(x) \approx & 2 [ \Phi'(x)^2 + \tfrac{1}{2} \Phi''(x) ( \Phi(x) - \Phi(a_0) )  ] \, f_\Phi (a_0,x,a_2) 
\nonumber\\
& + \Phi''(x) f_\Phi(a_0,x) \label{mf2} .
\end{align}
The differentiation formulas Eqs.~\refe{mf1}, \refe{mf2} are exact for polynomials up to degree
one and two, respectively, in the variable $y=\Phi(x)$.

\begin{table*}[t]\label{tab1}
\vspace*{-2em}
\centering
 \begin{tabular}{lllll}
Discretization & Grid & $N$ & $h$ & Numerical result / exact \\[1pt]\hline
\noalign{\vskip 5pt}    
Eq. \refe{f2uniform} & uniform
& 1000 	& 0.005 & 1.128179 \\
&& 10000	& 0.0005 & 1.099101 \\
&& 100000	& 0.00005 & 1.080742 \\
&& 1000000	& 0.000005 & 1.068109 \\
Eq. \refe{f2} & non-uniform 
&  1000 	&  & 1.060085 \\
&& 10000	&  & 1.047077 \\
&& 100000	&  & 1.038694 \\
&& 1000000	&  & 1.032845 \\
Eq. \refe{mf2} with $\Phi(x)=\sqrt{x}$ & uniform 
& 100	& 0.05	& 0.998358 \\
&& 500	& 0.01 & 0.999920 \\
 \end{tabular}
\caption{Comparison of different numerical methods for the eigenvalue problem of the two-dimensional
Schr\"odinger equation, for the exactly solvable case of the two-dimensional harmonic oscillator.
The problem is discretized in the range $0 < r < x_m\!=\!5$ on
a grid with $N$ points. For a uniform grid the spacing is $h$, the non-uniform grid points
are taken to be $n^2/(N^2 x_m)$, $n=0\ldots N$. The numerical result  
shown in the last column is the lowest eigenvalue (ground state) $E$, divided by the exact value $2$.}
\end{table*}

\section{Application: Eigenvalue problem for the two-dimensional Schr\"odinger equation}
\label{sec-4}

As an application of the formulas of Sec. \ref{sec-3}, we consider the eigenvalue problem
for the two-dimensional Schr\"odinger equation.
The time-independent Schr\"odinger equation in two dimensions reads
$(- \nabla^2 + V(\vec{r})) \Psi(\vec{r}) = E \Psi(\vec{r})$, and we assume that the potential
is radially symmetric $V=V(|\vec{r}|)$.
The problem is to find the energies (eigenvalues) $E$ and the wave functions (eigenvectors) 
$\Psi(\vec{r})$. Using polar coordinates $(r,\phi)$ and writing
$\Psi(\vec{r}) = \sum_{\ell=-\infty}^\infty e^{i \ell \phi} \psi(r)/\sqrt{r}$, the problem is reduced
to a radial equation, an ordinary differential equation. Here we consider only the equation for $\ell=0$,
it reads
\begin{equation}\label{schro2D}
\Bigl \{ -\frac{d^2}{dr^2} - \frac{1/4}{r^2}+ V(r) \Bigr\} \psi(r) = E \psi(r) .
\end{equation}
The boundary conditions for this eigenvalue problem for bound states are
\begin{equation}\label{bc}
\psi(0) =0, \qquad
\psi(\infty)=0.
\end{equation}

The straightforward method to solve this eigenvalue problem numerically is 
to discretize the equation \refe{schro2D} on a grid and solve
the resulting matrix eigenvalue problem. The discretization of the second derivative 
can be done using Eq. \refe{f2}.
It is found, however, that this method produces very inaccurate results. The convergence
of the numerical values of the eigenvalues is very slow as the number of grid points is increased,
both for a uniform and a non-uniform grid (see Table \ref{tab1}). 
Also, one sees that the wave function obtained using this method is inaccurate
for points close to the origin. The reason for this behavior is that the exact solution of Eq.~\refe{schro2D}
behaves for small $r$ as
\begin{equation}\label{psir}
 \psi(r) \sim \sqrt{r} .
\end{equation}
This leads to large errors in the discretization formula of Eq.~\refe{f2}, since the error is proportional to
the derivatives of the function [cf.\ Eq. \refe{NFE2}] which are here very large for small $r$ 
[cf.\ Eq.~\refe{psir}].

The discretization formula derived in Sec.\ \ref{sec-3} can be used for this problem to obtain accurate results.
One may use Eq.~\refe{mf2} with $\Phi(x)=\sqrt{x}$ to obtain a discretization for the second derivative, 
and use it to solve the eigenvalue problem for the two-dimensional Schr\"odinger equation~\refe{schro2D}.

To assess the accuracy of the numerical methods, we compare the results of the numerical
calculation of eigenvalues with an exact solution. For $V(r)=r^2$, the eigenvalue problem 
of Eq. \refe{schro2D} can be solved exactly (this is the harmonic oscillator in two
dimensions). The lowest eigenvalue is $E=2$ (in the units used here) and the corresponding eigenfunction is $\psi(r)=\sqrt{r}e^{-r^2/2}$.
Table \ref{tab1} compares the results for this problem obtained with different discretization schemes.
We compare the standard discretization on a uniform grid [Eq.~\refe{f2uniform}]
and on a non-uniform grid [Eq.~\refe{f2}] with
the results obtained by discretizing using modified divided differences, Eq. \refe{mf2}, with $\Phi(x)=\sqrt{x}$.
It is seen that the discretization formula with modified divided differences
is clearly superior to the two other methods.

Finally, to be fair it should be noted that the eigenvalue problem for two-dimensional 
$s$-wave Schr\"odinger equation can also
be treated with other methods. A substitution $\phi(r)=\psi(r)/\sqrt{r}$ brings 
Eq.~\refe{schro2D} to a form where the wave function is better behaved at the origin (the boundary condition of
Eq.~\refe{bc} has to be modified), and then the standard discretization procedure, even on
 a uniform grid, produces accurate results.

\section{Conclusions}

A method of discretizing derivatives of a function is presented which derives from the
notion of modified divided differences with respect to an increasing function $\Phi(x)$. 
The resulting formulas for derivatives correspond to polynomial interpolation 
in the variable $y=\Phi(x)$. The expressions generalize a formula due to Zadorin~\cite{Zadorin}.
The method is found to be useful for the eigenvalue problem for the two-dimensional Schr\"odinger 
equation, which describes diverse physical systems such as 
solid-state hetersotructures~\cite{Smith,Reimann}, layered superconductors~\cite{Chen},
as well as cold atoms in restricted geometries~\cite{Baranov}.
For an exactly solvable case, the method is compared to the usual discretization formulas 
and found to be more accurate by orders of magnitude.

%

\end{document}